\documentclass{emulateapj}
\def \emulmacro {}
\usepackage{graphicx}
\usepackage{amsmath}
\usepackage{acronym}
\newcommand{\hefourlong}{\mbox{HE\,0435$-$1223}}
\newcommand{\hefour}{\mbox{HE\,$0435$}}
\newcommand{\qtwotwo}{\mbox{Q\,$2237$}}
\newcommand{\qtwotwolong}{\mbox{Q\,2237$+$0305}}
\newcommand{\cxo}{the \textit{Chandra X-ray Observatory}}
\newcommand{\chandra}{\textit{Chandra}}
\newcommand{\err}[2]{\ensuremath{^{_{+#1}}_{^{-#2}}}}
\newacro{SMARTS}[SMARTS]{Small and Moderate Aperture Research Telescope System}
\newacro{ANDICAM}[ANDICAM]{A Novel Double Imaging Camera}
\newacro{UV}[UV]{ultraviolet}
\newacro{IR}[IR]{infrared}
\newacro{HST}[\textit{HST}]{\textit{Hubble Space Telescope}}
\newacro{P3M}[P3M]{particle-particle/particle-mesh}
\newacro{NFW}[NFW]{Navarro-Frenk-White}
\newacro{CMB}[CMB]{cosmic microwave background}
\newacro{PSF}[PSF]{point spread function}
\newacro{FWHM}[FWHM]{full width at half maximum}
\newacro{WCS}[WCS]{world coordinate system}
\newacro{SIS}[SIS]{singular isothermal sphere}
\newacro{WFC3}[WFC3]{the Wide Field Camera 3}
\begin{document}
%%%%%%%%%%%%%%%%%%%%%%%%%%%%%%%%%%%%%%%%%%%%%%
\bibliographystyle{apj}
%%%%%%%%%%%%%%%%%%%%%%%%%%%%%%%%%%%%%%%%%%%%%%
\shorttitle{OPTICAL, UV, \& X-RAY STRUCTURE OF HE\,0435$-$1223}
\shortauthors{BLACKBURNE ET AL.}
\slugcomment{ }
%%%%%%%%%%%%%%%%%%%%%%%%%%%%%%%%%%%%%%%%%%%%%% 
\title{The Optical, Ultraviolet, and X-ray Structure of the Quasar HE\,0435$-$1223\altaffilmark{1}}

\author{
  Jeffrey~A.~Blackburne\altaffilmark{2},
  Christopher~S.~Kochanek\altaffilmark{2,3},
  Bin~Chen\altaffilmark{4},
  Xinyu~Dai\altaffilmark{4}, \&
  George~Chartas\altaffilmark{5}
}

\altaffiltext{1}{Based on observations made with the NASA/ESA Hubble
Space Telescope, obtained at the Space Telescope Science Institute,
which is operated by the Association of Universities for Research in
Astronomy, Inc., under NASA contract NAS 5-26555. These observations
are associated with programs \#11732 and \#12324.}
\altaffiltext{2}{Department of Astronomy, The Ohio State University,
  140 West 18th Avenue, Columbus, OH 43210, USA;
  blackburne@astronomy.ohio-state.edu}
\altaffiltext{3}{Center for Cosmology and AstroParticle Physics, The
  Ohio State University, Columbus, OH 43210,
  USA}
\altaffiltext{4}{Homer L. Dodge Department of Physics and Astronomy,
  University of Oklahoma, Norman, OK 73019, USA}
\altaffiltext{5}{Department of Physics and Astronomy, College of
  Charleston, Charleston, SC 29424, USA}
\addtocounter{footnote}{5}
%%%%%%%%%%%%%%%%%%%%%%%%%%%%%%%%%%%%%%%%%%%%%%
\begin{abstract}

Microlensing has proved an effective probe of the structure of the
innermost regions of quasars, and an important test of accretion disk
models. We present light curves of the lensed quasar \hefourlong\ in
the $R$ band and in the ultraviolet, and consider them together with
X-ray light curves in two energy bands that are presented in a
companion paper. Using a Bayesian Monte Carlo method, we constrain the
size of the accretion disk in the rest-frame near- and far-UV, and
constrain for the first time the size of the X-ray emission regions in
two X-ray energy bands. The $R$-band scale size of the accretion disk
is about $10^{15.23}$\,cm ($\sim$$23r_g$), slightly smaller than
previous estimates, but larger than would be predicted from the quasar
flux. In the UV, the source size is weakly constrained, with a strong
prior dependence. The UV to $R$-band size ratio is consistent with the
thin disk model prediction, with large error bars. In soft and hard
X-rays, the source size is smaller than $\sim$$10^{14.8}$\,cm
($\sim$$10r_g$) at 95\% confidence. We do not find evidence of
structure in the X-ray emission region, as the most likely value for
the ratio of the hard X-ray size to the soft X-ray size is unity.
Finally, we find that the most likely value for the mean mass of stars
in the lens galaxy is $\sim$$0.3M_\odot$, consistent with other
studies.

\end{abstract}

%%%%%%%%%%%%%%%%%%%%%%%%%%%%%%%%%%%%%%%%%%%%%%

\section{Introduction}
\label{sec:intro}

The structure of the innermost regions of quasars and active galactic
nuclei, where the X-ray and continuum optical and ultraviolet emission
originates, is not well understood. The thin accretion disk model
\citep{Shakura:1973p337, Novikov:1973p343} is widely used to explain
the optical and \ac{UV} continuum, but has had difficulty reproducing
its observed spectral shape \citep[e.g.,][]{Blaes:2001p560}. Some
progress has been made using quasar variability
\citep[e.g.,][]{Kelly:2009p895, MacLeod:2010p1014, Dexter:2011pL24}
and ``disk reverberation'' \citep[e.g.,][]{Collier:2001p1527}, but the
issue remains unclear. The structure of the X-ray emitting regions is
even more mysterious, since the maximum temperature at the inner disk
edge is too low to produce X-rays. Instead, a jet or a hot corona
above the disk have been suggested as the X-ray source, giving rise to
direct and disk-reflected spectral components \citep[see, e.g., the
review by][]{Reynolds:2003p389}. Decisive tests of these models have
proved to be elusive.

Gravitational microlensing of strongly lensed quasars has emerged as
an effective technique for measuring the spatial structure of quasar
accretion disks. The stars in the foreground lens galaxy magnify (or
demagnify) the multiple quasar images, causing flux variability that
is not correlated as would be expected from intrinsic quasar
variability. The ratio of the angular size of the emission region to
the Einstein radius of a star in the lens galaxy determines the
strength of the variability, with large source sizes smoothing out the
light curves. Since the accretion disk temperature falls with radius,
the source is larger at longer wavelengths, and the microlensing
effect is weaker. For a standard thin disk, where the effective
temperature scales as the $\beta = 3/4$ power of the radius, the
characteristic size of the disk varies with wavelength as
$\lambda^{1/\beta}$. Several studies have used microlensing to put
limits on accretion disk sizes, generally finding that they are larger
at fixed wavelength than predicted by the thin disk model, and much
larger than sizes calculated using the optical flux
\citep{Pooley:2007p19, Anguita:2008p327, Morgan:2010p1129}. In
particular, \citet{Morgan:2010p1129} find that disk sizes scale with
black hole mass in a manner consistent with the $M_\mathrm{BH}^{2/3}$
scaling expected for a roughly constant Eddington fraction. There are
also studies of the dependence of the accretion disk size on
wavelength, which generally find that the size increases with
wavelength \citep{Poindexter:2008p34, Bate:2008p1955,
Eigenbrod:2008p933, Floyd:2009p233, Blackburne:2011p34,
Mosquera:2011p145}. One means of solving the size discrepancy is to
have a flatter temperature profile, more like the $T \propto R^{1/2}$
profile of an irradiated disk \citep[see][]{Morgan:2010p1129}, but so
far the uncertainties in the slope estimates from these studies are
too large to distinguish between these models. It does appear that
scattering on large scales and contamination from broad line emission
have too small an effect to resolve the problem \citep{Dai:2010p278,
Morgan:2010p1129, Mosquera:2011p145}.

In general, the observed effects of microlensing are stronger in
X-rays than at optical wavelengths, indicating that the X-rays are
emitted from a very compact region
\citep[e.g.,][]{Pooley:2007p19,Morgan:2008p755, Chartas:2009p174,
Dai:2010p278}.  A smaller source size also implies a shorter
source crossing time, so microlensing variability time scales
will tend to be shorter for the X-ray emission (e.g., the
simulations of \citep{Jovanovic2008}). 
Recently \citet{Chen:2011pL34} demonstrated that the
hard X-rays from the quasar \qtwotwolong\ are even more strongly
microlensed than the soft X-rays, indicating that there may be
temperature structure within the electron gas that presumably
generates the X-ray emission.

At a redshift $z_S = 1.689$, \hefourlong\ \citep[][hereafter
\hefour]{Wisotzki:2002p17, Morgan:2005p2531} is lensed by a foreground
early-type galaxy at a redshift of $z_L = 0.46$ into four images
arranged in a nearly perfect cross configuration. An $H$-band image of
the lens system is given in Figure~1 of \citet{Kochanek:2006p47}. In
this paper, we present light curves of the four images of the quasar
in the $R$ band and in the \ac{UV}, and additionally use light curves
in two X-ray energy bands from a companion paper, \cite{Chen2012}.
  We use these light curves to constrain the size of
the source at all four wavelengths. The whole is greater than the sum
of the parts: the simultaneous analysis of all four wavelengths yields
better constraints than any separate ``monochromatic'' analysis. In
addition, combining the \ac{UV} emission, which is thought to
originate in the inner regions of the accretion disk, with the optical
emission from larger radii gives us leverage to probe the change in
size of the disk with wavelength. Likewise, because we have both hard
and soft X-ray light curves, we can for the first time put a
quantitative constraint on the ratio of sizes in the two energy
bands. Finally, we put weak constraints on the mean mass of the stars
in the lensing galaxy. We outline the data in Section~\ref{sec:data},
describe our analysis method in Section~\ref{sec:method}, discuss the
results in Section~\ref{sec:results}, and briefly conclude in
Section~\ref{sec:conclusions}.

%%%%%%%%%%%%%%%%%%%%%%%%%%%%%%%%%%%%%%%%%%%%%%
\section{Data}
\label{sec:data}

\ifx \emulmacro \undefined
\else
\begin{figure*}
  \includegraphics[width=\textwidth]{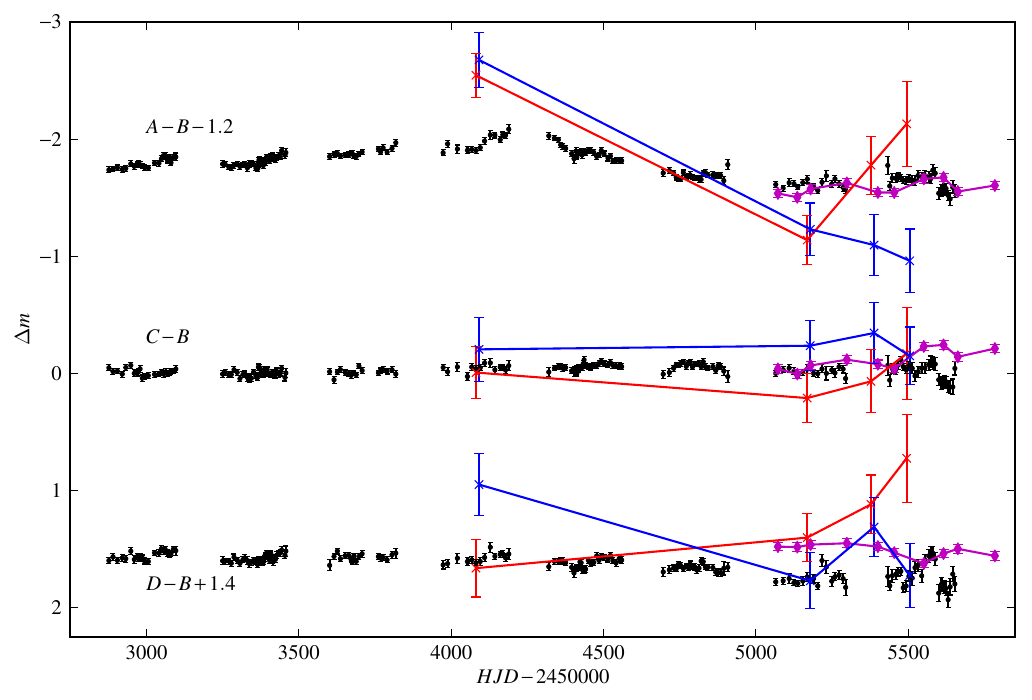}
  \caption{Light curves of the flux ratios between the images of
  \hefour, expressed in magnitudes. Black circles, magenta diamonds,
  red crosses, and blue crosses show the $R$-band, UV, soft X-ray, and
  hard X-ray data, respectively. The optical and UV curves have been
  corrected for the lensing time delays as described in
  \protect{Section~\ref{sec:data}}, and the optical light curve has
  been binned by a factor of two. Any remaining variability is
  considered to be due to microlensing. The light curves as shown are
  the input to the microlensing simulation. It is clear that the X-ray
  light curves have significantly more microlensing variability than
  the optical and UV curves.}
  \label{fig:lc}
\end{figure*}

\fi

We use eight seasons of optical $R$-band photometry resulting from an
ongoing monitoring campaign using A Novel Double Imaging Camera
\citep[ANDICAM,~][]{DePoy:2003p827} on the \ac{SMARTS} 1.3-m
telescope. This well-sampled light curve is supplemented by 10 epochs
of \ac{UV} data from \ac{WFC3} aboard the \ac{HST} and 4 epochs of
X-ray data from \cxo. In the rest frame of the quasar, these
wavelengths correspond to the near and far \ac{UV}, and to harder
X-rays, but for convenience we usually refer to them according to their
observed-frame bands.

\subsection{Optical data}
\label{sec:optdata}

The magnitudes of the four images of \hefour\ have been monitored
since 2003 August in the $R$ band (rest frame 2417\,\AA), with a
cadence varying from one day to about a week. An automated pipeline
subtracts instrumental bias and flat-fields the images in the standard
way. Three to six images are obtained per epoch, and we combine them
to increase the signal to noise ratio and reject cosmic rays. We
measure the fluxes of the quasar images and the lens galaxy using the
\ac{PSF} fitting method described in detail by
\citet{Kochanek:2006p47}. The width of the \ac{PSF} is measured using
the SExtractor software package \citep{Bertin:1996p393}, and we reject
epochs where the \ac{FWHM} is greater than $2\farcs{}5$ or where the
sky background is particularly high due to cloud cover or excessive
moonlight. For the handful of remaining epochs where the magnitudes
look grossly anomalous compared to neighboring epochs, we re-examine
the imaging data and invariably find that an uncorrected cosmic ray
has collided with one of the quasar images. In these cases we mask the
cosmic ray-affected pixels by hand and repeat the image combination
and flux measurements. The light curves are given in
Table~\ref{tab:optlc}. The $R$ band covers a region of the redshifted
quasar spectrum free of strong emission lines, but there may be
contamination from Fe\textsc{ii} emission and the Balmer
continuum. However, the equivalent width is small enough that it
should not strongly influence our results \citep[see][]{Dai:2010p278}.

We shift the light curves to correct for the light travel time delay
between the quasar images. We use delay values $\Delta t_\mathrm{BA} =
9.5$ days, $\Delta t_\mathrm{BC} = 7.2$ days, and $\Delta
t_\mathrm{BD} = -6.6$ days. These values are the result of fits to the
full optical light curves using the method of
\citet{Kochanek:2006p47}. Though we do not exhaustively explore the
range of parameters used for these fits, the resulting values are
consistent with the values of \citet{Courbin:2011p53}, and the
intrinsic variability of the quasar is not strong enough for small
($<1$ day) errors in the time delay corrections to have noticeable
effects on our microlensing analysis. After being shifted, the light
curves of images A, C, and D are resampled at the epochs of image B's
light curve using linear interpolation within each individual
season. We do not interpolate across seasonal gaps. Finally, we bin
each season of the light curves to reduce the total number of epochs
by a factor of 2. This helps to reduce the computational cost of our
microlensing analysis, and does not significantly degrade our
sensitivity to microlensing because its time scales are still longer
than the sampling of the light curves. The delay-corrected flux ratios
resulting from this processing are shown in
Figure~\ref{fig:lc}. Taking the ratio cancels out the intrinsic
variability of the quasar, leaving only microlensing variability. At
this wavelength, the microlensing variability is not dramatic, but can
be seen, particularly in the $A-B$ curve between the fourth and fifth
seasons.

\subsection{UV data}
\label{sec:uvdata}

\hefour\ was observed at 10 epochs between 2009 August and 2011 August
using \ac{WFC3} aboard the \ac{HST}. The observations used the F275W
filter in the UVIS channel, providing flux ratios at
2750\,\AA\ (1023\,\AA\ in the source rest frame). At each epoch there
are four exposures of 630\,s apiece, for a total of 2520\,s per
epoch. For one epoch, two of the four exposures were lost to a shutter
failure, but we are able to use the remaining two exposures. We use
the Multidrizzle task within the PyRAF software package\footnote{PyRAF
  and Multidrizzle are products of the Space Telescope Science
  Institute, which is operated by AURA for NASA.} to combine the four
flat-fielded images and reject cosmic rays. The \acp{WCS} provided
with the images enables the software to accurately align the
images. Examining the bad pixel masks produced by this process, we
find that several pixels at the center of each quasar image are
incorrectly marked as cosmic rays, so we manually edit the masks to
mark these pixels as good, and repeat the final drizzle step. The
result is a single clean and accurate image for each epoch. At this
wavelength, the images are remarkably empty, with few sources
approaching the brightness of the lensed quasar. In particular, the
lens galaxy and quasar host galaxy are not detected. 

\ifx \emulmacro \undefined
\else
\begin{deluxetable*}{ccccc}
\tablewidth{0pt}
\tablecaption{$R$-band Light Curves
  \label{tab:optlc}}
\tablehead{
  \colhead{$\mathrm{HJD}-2450000$} &
  \colhead{A} &
  \colhead{B} &
  \colhead{C} &
  \colhead{D}
} 
\startdata
$2863.883$ & $2.065\pm0.012$ & $2.632\pm0.019$ & $2.592\pm0.018$ & $2.821\pm0.022$ \\ 
$2871.829$ & $2.140\pm0.014$ & $2.628\pm0.022$ & $2.561\pm0.019$ & $2.790\pm0.023$ \\ 
$2877.843$ & $2.134\pm0.012$ & $2.644\pm0.017$ & $2.670\pm0.016$ & $2.803\pm0.018$ \\ 
$2884.784$ & $2.141\pm0.012$ & $2.653\pm0.017$ & $2.661\pm0.017$ & $2.870\pm0.019$ \\ 
$2891.827$ & $2.172\pm0.015$ & $2.689\pm0.023$ & $2.685\pm0.022$ & $2.831\pm0.024$ \\ 
$2899.843$ & $2.171\pm0.012$ & $2.728\pm0.018$ & $2.712\pm0.017$ & $2.856\pm0.019$ \\ 
$2906.848$ & $2.184\pm0.008$ & $2.729\pm0.011$ & $2.745\pm0.011$ & $2.891\pm0.012$ \\ 
$2916.803$ & $2.192\pm0.010$ & $2.720\pm0.015$ & $2.713\pm0.014$ & $2.953\pm0.016$ \\ 
$2919.828$ & $2.177\pm0.013$ & $2.759\pm0.020$ & $2.738\pm0.019$ & $2.910\pm0.021$ \\ 
$2926.780$ & $2.186\pm0.012$ & $2.736\pm0.018$ & $2.643\pm0.016$ & $2.885\pm0.019$
\enddata
\tablecomments{~Light curves are in uncalibrated
magnitudes. This table is published in its entirety online. A portion
is shown here for guidance regarding its form and content.}
\end{deluxetable*}

\begin{deluxetable*}{ccccc}
\tablewidth{0pt}
\tablecaption{Ultraviolet (F275W) Light Curves
  \label{tab:uvlc}}
\tablehead{
  \colhead{$\mathrm{HJD}-2450000$} &
  \colhead{A} &
  \colhead{B} &
  \colhead{C} &
  \colhead{D}
} 
\startdata
$5072.079$ & $18.566\pm0.002$ & $18.917\pm0.003$ & $18.867\pm0.003$ & $19.005\pm0.003$ \\
$5135.646$ & $18.664\pm0.003$ & $19.004\pm0.003$ & $18.983\pm0.003$ & $19.096\pm0.003$ \\
$5179.185$ & $18.830\pm0.004$ & $19.198\pm0.005$ & $19.146\pm0.005$ & $19.288\pm0.006$ \\
$5298.162$ & $18.780\pm0.003$ & $19.197\pm0.003$ & $19.090\pm0.003$ & $19.247\pm0.004$ \\
$5400.013$ & $18.684\pm0.003$ & $19.064\pm0.003$ & $18.953\pm0.003$ & $19.136\pm0.003$ \\
$5453.678$ & $18.889\pm0.003$ & $19.244\pm0.004$ & $19.204\pm0.003$ & $19.408\pm0.004$ \\
$5550.662$ & $18.992\pm0.003$ & $19.433\pm0.004$ & $19.219\pm0.003$ & $19.676\pm0.004$ \\
$5615.296$ & $18.878\pm0.003$ & $19.323\pm0.004$ & $19.104\pm0.003$ & $19.443\pm0.004$ \\
$5662.224$ & $18.742\pm0.003$ & $19.066\pm0.003$ & $18.949\pm0.003$ & $19.119\pm0.003$ \\
$5783.750$ & $18.450\pm0.003$ & $18.854\pm0.003$ & $18.642\pm0.003$ & $19.012\pm0.004$
\enddata
\tablecomments{~Light curves are in ST magnitudes.}
\end{deluxetable*}

\fi

We use DoPHOT \citep{Schechter:1993p1342} to perform aperture
photometry on the quasar components in each epoch, using an aperture
0\farcs85 on a side, wide enough to capture essentially all of the
flux in a given quasar component but small enough that the other
components do not contaminate the photometry. We convert the
instrumental fluxes into ST magnitudes using the standard header
keywords PHOTFLAM and PHOTZPT. In this filter, the offset from ST to
AB magnitudes is $m_\mathrm{AB} - m_\mathrm{ST} = 1.532$. The ST
magnitudes of the four quasar components for all 10 epochs are listed
with their formal uncertainties in Table~\ref{tab:uvlc}, and the
magnitude differences are plotted in Figure~\ref{fig:lc}. The
magnitudes in Table~\ref{tab:uvlc} are uncorrected for Galactic
extinction, which we calculate to be $0.36$ magnitudes along this
line of sight, using the \citet{Schlegel:1998p525} value and the
\citet{Cardelli:1989p245} $R_V=3.1$ extinction law. Since we are only
concerned with magnitude differences, this overall offset has no
effect on our calculations. At the redshift of this quasar, Ly$\beta$
\textsc{Ovi}, and other emission lines overlap with the F275W filter,
but the passband is wide enough that the continuum dominates the flux;
we estimate that about 10\% of the total flux is in the emission
lines. Contamination from the spatially extended emission line region
may bias our results toward larger \ac{UV} source sizes, but the
effect should not be a strong one \citep{Dai:2010p278}. Because the
latest epoch of \ac{UV} data extends beyond our optical light curve,
and because our simulations were begun before it was collected, we use
only the first 9 epochs of the \ac{UV} light curves for our
microlensing simulations.

We shift the \ac{UV} light curves of quasar images A, C, and D
according to the time delays given in Section~\ref{sec:optdata}, and
use linear interpolation to resample them contemporaneously with
B. The time delays between the quasar images are small compared to the
typical separation between epochs, so the adjustments are small. But
there is potential for systematic error because of intrinsic quasar
variability that is not well-described by our linear interpolation, so
we broaden the uncertainties on the light curves to account for
this. To estimate this extra uncertainty, we scale and shift the
\emph{optical} light curves so that they match as closely as possible
the \ac{UV} light curves, and treat them as an estimate of the \ac{UV}
variability. The typical absolute difference between the interpolated
\ac{UV} magnitudes and the scaled and shifted optical light curve is
$\sim$0.036 mags, and we add this in quadrature to the formal
uncertainties. Since the statistical uncertainties are so small, this
systematic contribution dominates the error budget for the \ac{UV}
light curves.

The lack of flux from the lens galaxy in the \ac{UV} affords us an
opportunity to investigate the existence of a fifth image of the
quasar. Lensing theory generically predicts an odd number of quasar
images, but in practice lensed quasars almost always have either 2 or
4 observed images. The ``missing image'' is located at the maximum of
the light travel time near the center of the lens galaxy, and is
strongly demagnified due to the sharp curvature of the potential
\citep{Rusin:2001pL33, Mao:2001p301, Keeton:2003p17,
Rusin:2005pL93}. This makes central images an interesting probe of
both the innermost density profiles and central black holes of distant
galaxies, but so far only two central images have been unambiguously
detected \citep{Winn:2004p613, Inada:2005pL7, Inada:2008pL27}. To
check for a central image at \ac{UV} wavelengths, we create a deep
image of \hefour\ by stacking the frames from every epoch using
Multidrizzle. There are small ($\sim$$1''$) offsets between the
\acp{WCS} of each epoch, so we measure the pixel shifts between the
single-epoch images using cross-correlation and provide them to
Multidrizzle. We use \ac{PSF} fitting to carefully subtract the wings
of the four bright quasar images and test for a central image. The
\ac{PSF} model is constructed from the (exposure-weighted) sum of
oversampled and appropriately rotated Tiny Tim \citep{Krist:2011p0J}
model \acp{PSF} for the set of 10 epochs. We fix the relative
positions of the five quasar images, using the positions of the bright
four and of the lens galaxy from the \textsc{Castles} database
\citep{Falco:2001p25}, and allow the overall position offset and the
five normalizations to vary. The resulting flux ratios of images A
through D agree with our previous aperture measurements to about
2\%. This is not as good as the $\sim$0.005-mag formal errors in
Table~\ref{tab:uvlc}, but this may be explained by the fact that the
central region of the model \ac{PSF} does not match that of the
real-life \ac{PSF} very well. We rule out a central image brighter
than 26.3 ST magnitudes at 95\% confidence; this implies a flux ratio
relative to the faintest image $f_5/f_D < 0.002$. This non-detection
probably arises from either a massive central black hole or a steep
density profile in the lens galaxy, but it is difficult to say
anything more quantitative from these \ac{UV} observations because of
the possibility of extinction within the lens galaxy. Likewise, if we
\emph{had} detected flux in the central regions, distinguishing quasar
emission from nuclear star formation or weak AGN activity in the lens
galaxy would have been difficult with these observations.

\subsection{X-ray data}
\label{sec:xraydata}

\hefour\ was observed on 2009/12/07, 2010/07/04, and 2010/10/29 using
\chandra/ACIS \citep{Weisskopf:2002p1, Garmire:2003p28}, with an
exposure time of 12.8\,ks for each observation. Including an archival
observation made in 2006 \citep{Blackburne:2011p34, Pooley:2012p111},
we have accumulated four \chandra\ observations of the quasar. The
details of the reduction of these observation are presented in a
companion paper (\citealt{Chen2012}). We separate the
photon events into soft (0.4\,--\,1.3\,keV) and hard
(1.3\,--\,8.0\,keV) energy bands and perform imaging and spectral
analyses of the four images of \hefour\ in these bands and in the full
energy range. In the spectral analysis, we fit the spectra using a
power law plus Gaussian emission line model, representing the X-ray
continuum and the \mbox{Fe\,K$\alpha$} emission line, both modified by
Galactic absorption and absorption at the lens redshift. In the
imaging analysis, we use \ac{PSF} fitting to determine the photon flux
for each image in each epoch and each band. After obtaining the best
fits from the imaging analysis, we correct for absorption to obtain
the intrinsic photon fluxes for each image. The resulting four-epoch
X-ray light curves for each quasar image in three energy bands are
presented by Chen~et~al.~(2012), and we use the soft and hard band
light curves in our simulations. The flux ratios, expressed as
magnitude differences, are shown in Figure~\ref{fig:lc}.

Since the X-ray light curves are even more sparsely sampled than the
\ac{UV} light curves, it seems inappropriate to use interpolation to
attempt a correction for the systematic errors caused by lensing time
delays combined with intrinsic quasar variability. Instead, we simply
leave the X-ray fluxes uncorrected, and broaden the uncertainties to
account for this error source. Unlike the \ac{UV} light curves, the
statistical uncertainties are much larger than the systematic errors
that we expect from time delays.

%%%%%%%%%%%%%%%%%%%%%%%%%%%%%%%%%%%%%%%%%%%%%%
\section{Method}
\label{sec:method}

To constrain the source properties (i.e., the size and shape of the
quasar emission regions) and lens properties (i.e., the mean mass of
stars and peculiar velocity), we use many simulated light
curves. These simulations are compared to the observed light curves
and evaluated using a chi-square estimator. Each simulated light
curve, which is associated with a particular vector in parameter
space, is thus assigned a likelihood. We add them together to create a
joint likelihood distribution, which we then combine with priors on
the parameters to generate posterior probability distributions for the
quantities of interest using Bayes' theorem. This Bayesian Monte Carlo
approach is laid out in detail by \citet{Kochanek:2004p58}, and we use
a version that is updated by \citet{Poindexter:2010p658} to include
the random motions of the microlens stars. 

We model the light curves for the four quasar components at each of
the four wavelengths as an independent intrinsic quasar light curve
$S$ modified by a time-invariant and achromatic macro-magnification
$\mu_i$ due to the overall gravitational potential of the lens galaxy,
by the magnification due to microlensing $\delta\mu_i$, and by a
time-invariant ``catch-all'' term $\Delta\mu_i$. This last term
represents several possible systematic effects, such as differential
extinction in the lens galaxy \citep[e.g.,][]{Falco:1999p617,
Mosquera:2011p145}, millilensing by dark matter substructure
\citep[e.g.,][]{Kochanek:2004p69, Fadely:2011p101}, systematic errors
in the macro-magnification $\mu_i$, or low levels of unrecognized
contamination of the quasar flux by the lens or host
galaxy. Expressing these quantities in magnitudes, this is
% i=component; j=time; k=wavelength
\begin{align}
  \label{eqn:model}
  m_i\left(t_j,\lambda_k\right) &= S\left(t_j,\lambda_k\right) + \mu_i
  + \Delta\mu_i\left(\lambda_k\right) +
  \delta\mu_i\left(t_j,A(\lambda_k)\right) \nonumber \\
  &= S\left(t_j,\lambda_k\right) + \mu_{i,\mathrm{tot}} ~,
\end{align}
where $i$ enumerates the four quasar components (A, B, C, and D), and
$t_j$ and $\lambda_k$ denote the set of epochs and wavelength in our
light curves. The function $A(\lambda_k)$ is the area of the
quasar emission profile at wavelength $\lambda_k$; it is through this
function that the microlensing term $\delta\mu_i$ depends on
wavelength.

\subsection{Microlensing Simulations}
\label{sec:magsim}

We simulate the microlensing parts of our model (i.e., the
$\delta\mu_i$ term) using microlensing magnification patterns created
using the particle-particle/particle-mesh
\citep[P3M,~][]{Hockney:1981} method detailed by
\citet{Kochanek:2004p58}. The patterns give the microlensing
magnification of each quasar image as a function of the source's
position. The patterns appear as a network of intertwined
high-magnification caustics separated by lagoons of lower
magnification. Like \citet{Poindexter:2010p668, Poindexter:2010p658},
we use dynamic patterns, meaning that they are recalculated for each
epoch. This allows us to take into account the random velocities of
the stellar microlenses as well as the parallactic motion of the
earth. The patterns are periodic by construction in both the lens and
source planes, allowing us to wrap light curves and stellar motions
across the edges, and eliminating edge effects when convolving the
patterns.

The general characteristics of the magnification patterns are
determined for each quasar image by the local shear $\gamma$ and the
local convergence $\kappa_\mathrm{tot} = \kappa_s + \kappa_*$, where
$\kappa_s$ is the convergence due to smoothly distributed matter and
$\kappa_*$ is due to the microlens stars. These quantities, as well as
the macro-magnifications $\mu_i$, are determined by modeling the
lensing of the system as a whole. We use models from
\citet{Kochanek:2006p47}, who parametrize the relative contributions
of stars and of the dark matter halo using $f_{M/L}$, the
normalization of the stellar component relative to its best-fit
normalization with no dark matter halo. The particular models we use
set the effective radius of the stellar component $R_e$ to
$0\farcs{}86$ and the break radius of the \ac{NFW} halo $r_c$ to
$10\farcs{}0$. We use three values of $f_{M/L}$: 0.1, 0.3, and
0.95. This allows us to explore a range of values for this parameter,
and for all of our results we marginalize over it. Although there are
reasons to prefer a low value of $f_{M/L}$ \citep[see,
e.g.,][]{Kochanek:2006p47}, we conservatively elect not to weight some
values higher than others. This choice has almost no effect on our
results. The values of $\gamma$, $\kappa_s$, and $\kappa_*$ resulting
from these models, which we use to construct the magnification
patterns, are listed in Table~\ref{tab:modparam}.

\ifx \emulmacro \undefined
\else
\begin{deluxetable}{ccccccc}
\tablewidth{0pt}
\tablecaption{Lens Model Parameters
  \label{tab:modparam}}
\tablehead{
  \colhead{$f_{M/L}$} &
  \colhead{Img} &
  \colhead{$\kappa_\mathrm{tot}$} &
  \colhead{$\kappa_*$} &
  \colhead{$\gamma$} &
  \colhead{$\phi_\gamma$\tablenotemark{a}} &
  \colhead{$\mu$\tablenotemark{b}}
}
\startdata
0.1                  & A & 0.662 & 0.0151    & 0.223    & \phn63\fdg2 & $+15.6$ \\
                     & B & 0.789 & 0.0292    & 0.322    &    344\fdg7 & $-16.7$ \\
                     & C & 0.663 & 0.0151    & 0.226    &    270\fdg3 & $+16.1$ \\
                     & D & 0.839 & 0.0367    & 0.342    &    167\fdg0 & $-10.9$ \\
0.3                  & A & 0.547 & 0.0464    & 0.300    & \phn63\fdg0 & $+8.68$ \\
                     & B & 0.674 & 0.0886    & 0.469    &    344\fdg7 & $-8.79$ \\
                     & C & 0.548 & 0.0465    & 0.304    &    270\fdg5 & $+8.92$ \\
                     & D & 0.723 & 0.0109    & 0.511    &    167\fdg1 & $-5.43$ \\
0.95                 & A & 0.157 & 0.148\phn & 0.553    & \phn62\fdg6 & $+2.48$ \\ 
                     & B & 0.299 & 0.289\phn & 0.961    &    344\fdg7 & $-2.31$ \\ 
                     & C & 0.158 & 0.147\phn & 0.562    &    270\fdg8 & $+2.54$ \\ 
                     & D & 0.358 & 0.347\phn & 1.07\phn &    167\fdg0 & $-1.36$ 
\enddata
\tablenotetext{a}{Measured in degrees East of North.}
\tablenotetext{b}{Sign indicates parity. Magnifications calculated using
\mbox{\protect{$\mu =
[(1-\kappa_\mathrm{tot}+\gamma)(1-\kappa_\mathrm{tot}-\gamma)]^{-1}$}}.}
\end{deluxetable}

\fi

We randomly scatter stars across the region near each quasar component
to create the microlens convergence $\kappa_*$. Their masses are drawn
from a $dN/dM\propto M^{-1.3}$ power-law mass function with a
dynamic range of 50 between its maximum and minimum masses. This a
reasonable match to stellar mass functions
\citep{Poindexter:2010p658}, and the details should have little effect
on our results \citep[e.g.,][]{Wyithe:2000p51}. We use a range of
values for the mean mass of the stars, with $\langle M/M_\odot \rangle
= 0.03$, $0.1$, $0.3$, $1$, $3$, and $10$. The magnification patterns'
outer dimensions are 20 times the Einstein radius $R_\mathrm{Ein}$ of
this mean mass, where
\begin{align}
  R_\mathrm{Ein} &= D_\mathrm{OS} \theta_\mathrm{Ein} =
  \left(\frac{4G\langle M \rangle}{c^2} \frac{D_\mathrm{OS}
  D_\mathrm{LS}}{D_\mathrm{OL}}\right)^{1/2}\nonumber \\
  &= (5.42\times10^{16}~\mathrm{cm}) \langle M/M_\odot \rangle^{1/2} ~,
\end{align}
and $(D_\mathrm{OL}, D_\mathrm{OS}, D_\mathrm{LS}) = (1203, 1746,
1093)$~Mpc are the angular diameter distances from observer to lens,
observer to source, and lens to source, respectively, for $\Omega_M =
0.3$, $\Omega_\Lambda = 0.7$, and $H_0 =
70$\,km\,s$^{-1}$\,Mpc$^{-1}$. The patterns used for the optical
wavelength simulations are 4096 pixels on a side, so that their pixel
size is approximately $2.6\times10^{14} \langle M/M_\odot
\rangle^{1/2}$~cm. Since the mass of the central black hole is $\sim
5\times10^8 M_\odot$ \citep[based on the \textsc{Civ} line
  width,][]{Peng:2006p616}, this pixel size is about $3.6 \langle
M/M_\odot \rangle^{1/2}$ gravitational radii. \cite{Greene2010} and
\cite{Assef2011} compared black hole mass estimates from the CIV and 
H$\beta$/H$\alpha$ emission lines for a significant sample of lenses (unfortunately, 
HE~0435 was not one of them), finding that they were mutually consistent
given the usual $\sim 0.3$-$0.5$~dex uncertainties of such estimates.   
Since the emission is
more compact at \ac{UV} and X-ray wavelengths, for these simulations
we create more detailed versions of the same patterns, doubling the
resolution to 8192 pixels. Care must be taken when recreating the
patterns at higher resolution, because of certain details of the
pattern creation algorithm. Specifically, the shear $\gamma$ is
adjusted slightly from the input value so that the periodic scale is
an integer number of pixels in both planes. When creating the
high-resolution patterns we intentionally calculate this adjustment as
if the pattern dimension were 4096 instead of 8192. This results in
the desired periodicity, and avoids small distortions of the new
pattern relative to the old.

For each quasar image, the microlensing variability is simulated by
the movement of the quasar across the pattern, as well as the
evolution of the pattern itself due to the random motions of the
stars. The latter is made possible by our dynamic patterns, where each
star's position is dependent on epoch. We use a one-dimensional
velocity dispersion of 255\,km\,s$^{-1}$ in the lens rest frame,
estimated from the Einstein radius of a \ac{SIS} model of the
lens. \citet{Treu:2006p662} have shown this to be a good estimator for
the stellar velocity dispersion, and our value is consistent with the
measurement of \citet{Courbin:2011p53}. We assume that the lens galaxy
has negligible bulk rotational velocity. The effective velocity of the
source across the pattern is the vector sum of the cosmological
peculiar velocities of the observer, the lens, and the source,
properly scaled for geometry and cosmological time dilation. We follow
the approach of \citet{Poindexter:2010p658}, taking for the observer's
motion the \ac{CMB} dipole velocity \citep[from][]{Hinshaw:2009p225}
projected perpendicular to the line of sight, combined with the
parallax due to the earth's orbit, and treating the remaining two
velocities as normally distributed random variables. For \hefour, the
projected \ac{CMB} velocity is \mbox{$(363, -56)$\,km\,s$^{-1}$} East
and North, respectively. Using the empirical interpolation of
\citet{Mosquera:2011p96}, we estimate that the rest-frame peculiar
velocity dispersions for the lens and the source are $\sigma_L = 277$
and $\sigma_S = 227$\,km\,s$^{-1}$, respectively. Cosmological time
dilation and geometric projection effects combine to suppress the
contribution of the source, so that the lens velocity dominates the
effective velocity of the source across the pattern. The dispersion of
the prior on the effective velocity, projected to the lens plane, is
given by\footnote{In their Equation~12,
\protect{\citet{Poindexter:2010p658}} neglect a factor of $(1+z_S)$ in
the scaling of $\sigma_S$, but this introduces only a negligible error
in their velocity prior.}
\begin{equation}
  \sigma_\mathrm{eff}^2 = \sigma_L^2 + \sigma_S^2
  \left(\frac{1+z_L}{1+z_S}\right)^2 
  \left(\frac{D_\mathrm{OL}}{D_\mathrm{OS}}\right)^2
  = \left(290\,\mathrm{km\,s}^{-1}\right)^2 ~.
\end{equation}

The finite size of the source is simulated by convolving the
magnification patterns with the source's emission profile. We use a
standard thin disk model that radiates as a multitemperature
blackbody, with an effective temperature that varies inversely with
the $\beta = 3/4$ power of the radius. We neglect the inner edge of
the disk. Although it is not guaranteed that this is the correct model
for the accretion disk, in practice the details of the source's radial
profile are not very important \citep{Mortonson:2005p594,
Congdon:2007p263}.  Based on these simulations and our prior
observational studies of microlensing variability, the best present 
means of studying the spatial structure of emission regions is
to estimate the size at different wavelengths rather than using
a more complex parametrization of the spatial structure.   
The results can then be compared to any more
complex model for the emissivity or the temperature profile 
(e.g. \citep{Popovic2006}, \citep{Chen2013}) by matching the
half-light radius predicted by the model to our observational
constraint.  

The wavelength-dependent area of the disk
$A(\lambda)$ is defined as the area within the contour where $k_B
T_\mathrm{eff} = hc/\lambda$. We vary this area in logarithmic
intervals of 0.2 dex, using an adaptive algorithm to explore the peak
of the likelihood function, and stopping when the source is too small
to be resolved by the magnification pattern pixels, when the
likelihood falls to a very small fraction of the peak value, or when
30 values of the area have been explored. Like
\citet{Poindexter:2010p668}, we also vary the inclination of the disk
$\cos i$ and the position angle of its major axis $\phi_a$. We use five
inclinations, with $\cos i$ ranging from 1.0 (face-on) to 0.2 (nearly
edge-on) in steps of 0.2. Similarly, we use 9 major axis position
angles, ranging from $0^\circ$ to $160^\circ$. The inclination and
major axis position angle do not vary with wavelength, but the area is
allowed to vary independently in our four bands.

\subsection{Bayesian Monte Carlo}
\label{sec:bayesmc}

A handful of parameters determine the exact shape of the microlensing
light curve: the source parameters ($A$, $\cos i$, and $\phi_a$), the
lens parameters ($\langle M \rangle$ and $f_{M/L}$), and a particular
starting point and effective velocity of the path across the
magnification pattern. As described in detail by
\citet{Kochanek:2004p58} and \citet{Poindexter:2010p658}, we use Monte
Carlo integration to evaluate the Bayesian integrals over the space of
these parameters. For each combination of source and lens parameters,
we evaluate $12500$ trial paths across the magnification patterns. Each
trial is evaluated using the chi-square statistic
\begin{equation}
  \label{eqn:chisq}
  \chi_k^2 = \sum_i \sum_j \left[ \frac{m_i\left(t_j,\lambda_k\right) -
    S\left(t_j,\lambda_k\right) - \mu_{i,\mathrm{tot}}}
      {\sigma_i\left(t_j,\lambda_k\right)} \right]^2 ~,
\end{equation}
where $m_i$ and $\sigma_i$ are the measured magnitudes and their
uncertainties, and $S$ and $\mu_{i,\mathrm{tot}}$ are defined in
Equation~(\ref{eqn:model}). The intrinsic quasar light curve
$S\left(t_j,\lambda_k\right)$ and the ``catch-all'' term $\Delta
\mu_i(\lambda_k)$ are determined for each trial by minimization of
$\chi_k^2$. Trials with chi-square values above a predefined cutoff
value are discarded, as their contribution to the likelihood is
negligible. We impose a Gaussian prior on $\Delta \mu_i(\lambda_k)$,
with a mean of 0~mags and a variance of $(0.1\,\mathrm{mag})^2$, by
adding a suitable term to the chi-square. We conservatively allow this
value to vary independently for each wavelength, since some of its
contributors are wavelength-dependent (e.g., flux contamination from
the lens galaxy at optical wavelengths, or differential
extinction/absorption by the interstellar medium of the lens). We are
similarly conservative in electing not to put a prior on the shape of
the intrinsic source light curve, though in principle we could do so,
giving a likelihood boost to trials yielding light curves with
statistical properties matching those of other quasars of similar
redshift and luminosity \citep[see, e.g.,][]{Kozlowski:2010p927,
MacLeod:2010p1014}.

We are interested in trials that provide simultaneous good fits to our
optical, \ac{UV}, and X-ray light curves. In principle, it is possible
to simultaneously evaluate $\chi^2$ for a given trial at all
wavelengths, but given the length of our light curves, particularly in
the optical, we find it less computationally demanding to treat the
four wavelengths separately. We first amass a collection of trials
that provide reasonable fits to our $R$-band data, and then re-simulate
each of these using an independently varied source area for the
\ac{UV} light curve. This results in a new set of trials that fit both
the optical and \ac{UV} data. There is a balance between adding and
removing trials, since each input trial generates a number of new
trials with a range of \ac{UV} source areas, but some of these are
rejected due to the $\chi^2$ cutoff. Each of the resulting trials has
the same parameters as the trial from which it was generated, and
additionally a value for the \ac{UV} source area and a \ac{UV}
chi-square value and $N_\mathrm{dof}$. We continue this process for
the soft and hard X-ray light curves, using the output of one
simulation as the input for the next, and obtaining in the end a final
set of trials with independent optical, \ac{UV}, and soft and hard
X-ray source areas.

\ifx \emulmacro \undefined
\else
\begin{figure*}
  \includegraphics[width=\textwidth]{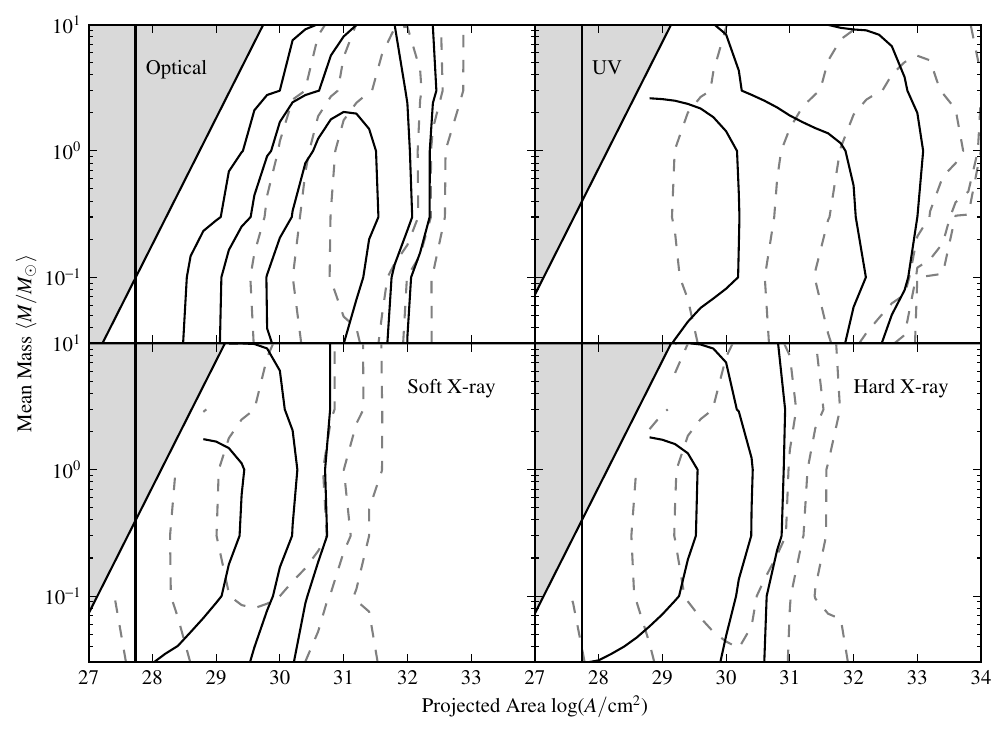}
  \caption{Joint posterior probability distributions for the projected
  area of the source and the microlens mean mass. The contours show
  39\%, 86\%, and 99\% enclosed probability (note the non-standard
  values). Black curves denote the logarithmic prior, and gray dashed
  curves the linear prior. The shaded zone indicates the region where
  the source area is smaller than the pixel size of the magnification
  patterns. The square of the black hole's gravitational radius (for
  $M_\mathrm{BH} = 5\times10^8M_\odot$) is shown with a vertical
  line. All other parameters are marginalized.}
  \label{fig:massarea}
\end{figure*}

\fi

We convert chi-square values to likelihoods using an incomplete Gamma
function, as derived by \citet{Kochanek:2004p58}. This function is
robust to small errors in the light curve uncertainties, even with
many degrees of freedom; in essence, it does not give ``extra credit''
to solutions with $\chi^2/N_\mathrm{dof} < 1$. We adapt it for
multiwavelength use, defining the likelihood as
\begin{equation}
\label{eqn:likelihood}
\mathcal{L} \propto \Gamma\left(\sum_k \frac{N_{\mathrm{dof},k}}{2}-1,
\sum_k \frac{\chi_k^2}{2f_{0,k}^2}\right) ~,
\end{equation}
where $\chi_k^2$ and $N_{\mathrm{dof},k}$ are the chi-square and
degrees of freedom at wavelength $\lambda_k$, and $f_{0,k}^2$ is the
factor by which we rescale the chi-square to account for errors in the
uncertainties. For the optical, \ac{UV}, soft X-ray, and hard X-ray
light curves, respectively, $N_{\mathrm{dof},k}$ is 585, 24, 9, and 9,
$f_{0,k}^2$ is 2.0, 1.75, 1.25 and 1.25, and the cutoff values of
$\chi_k^2$, above which trials are discarded, are 4.0, 3.5, 2.5, and
2.5.

We multiply the resulting likelihood distribution by appropriate
priors to generate posterior probability distributions for the
parameters of interest, marginalizing over the less-interesting
parameters. The prior is uniform for the starting position of the
trial paths, and normally distributed as described in
Section~\ref{sec:magsim} for their effective velocities. We experiment
with two other velocity priors: a broad (1000\,km\,s$^{-1}$
dispersion) Gaussian prior and a flat prior extending to $\pm
3000$\,km\,s$^{-1}$ in each direction. Such broad priors hamper our
ability to rule out large mean masses and large source sizes. With the
broad Gaussian prior, the peaks of the posterior probability
distributions for these variables by factors of roughly 3 toward
larger values (factors of roughly 10 for the source area). With the
flat velocity prior, the probability distribution for the mean mass
increases without bound for large values of $\langle M \rangle$,
leading to broad distributions for the source sizes, with almost no
upper limits. While it is accurate therefore to say that our results
are prior-dependent, it must be noted that these alternative priors
are not very physically likely. For the other parameters, we use
uniform logarithmic (e.g., $\langle M \rangle$, $f_{M/L}$) or linear
(e.g., $\cos i$, $\phi_a$) priors.

In particular, for the area $A$ and scale radius $r_s$ of the source
we explore the results of using both priors. Ideally, we would like to
have results that are independent of the priors, but single-epoch and
sparsely-sampled microlensing results tend to have some prior
dependence. In general, for a positive-definite scaled quantity of
uncertain scale, the standard prior is logarithmic so that all scales
are weighted equally, with $P(r_s) \propto r_s^{-1}$.  Clearly,
however, this prior becomes incorrect close to the gravitational
radius $r_g$ because we have a rough knowledge of the black hole mass
and the size of the emission region cannot be logarithmically smaller
than $r_g$.  Hence, on smaller scales near $r_g$ a linear prior with
$P(r_s)$ constant, or a linear prior cut off at some fraction of $r_g$
is more appropriate than a logarithmic prior. A possible compromise
that incorporates both of these priors would be to use a hybrid prior
$P(r_s) \propto (r_g + r_s)^{-1}$ which correctly switches between our
prior knowledge that the emission is unlikely to be logarithmically
smaller than $r_g$ but could be logarithmically larger. For clarity
and ease of comparison with previous results, we do not quote the
results of using this hybrid prior or show it in our figures, but it
may be imagined quite simply by inspection of the posterior
probability distributions that we do show: its distribution has the
shape of the linear-prior distribution at sizes smaller than the
gravitational radius $r_g$, smoothly transitioning to that of the
logarithmic-prior distribution at larger sizes. Since the pure
logarithmic prior makes the weakest claims about ruling out small
source sizes, we prefer this prior, and all the results that we give
concerning the area or size of the accretion use this prior, unless
stated otherwise.

%%%%%%%%%%%%%%%%%%%%%%%%%%%%%%%%%%%%%%%%%%%%%%
\section{Results and Discussion}
\label{sec:results}

\ifx \emulmacro \undefined
\else
\begin{figure*}
  \includegraphics[width=\textwidth]{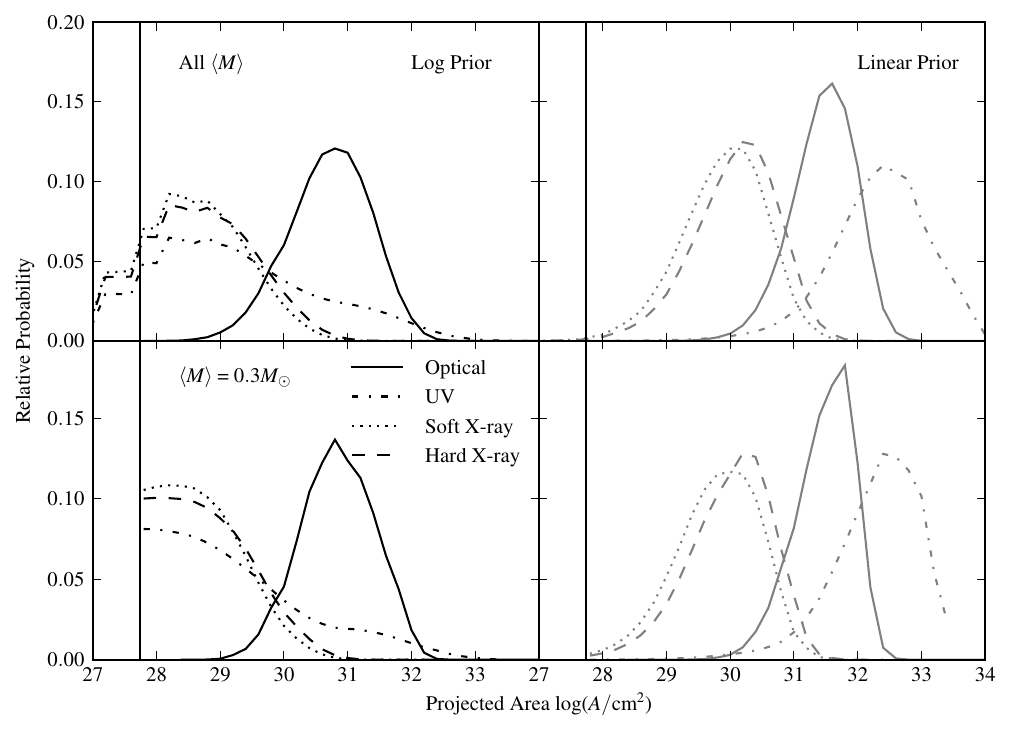}
  \caption{Posterior probability distributions for the projected area
  of the region producing the observed-frame $R$ band, UV, soft X-ray,
  and hard X-ray emission. The left (right) two panels use a
  logarithmic (linear) prior. In the top panels, all other variables
  have been marginalized, including $\langle M \rangle$, leading to an
  artificial decrease in probability at small areas (see
  \protect{Figure~\ref{fig:massarea}}). In the bottom panels, $\langle
  M \rangle$ is fixed at $0.3 M_\odot$. In all panels, the square of
  the black hole's gravitational radius is marked with a vertical
  line.}
  \label{fig:area}
\end{figure*}

\fi

\subsection{Accretion Disk Area}

Figures~\ref{fig:massarea} and \ref{fig:area} show posterior
probability distributions for the projected area of the quasar. In
Figure~\ref{fig:massarea} we plot the joint distribution of the source
area and the mean microlens mass at optical and \ac{UV} wavelengths
and in both soft and hard X-ray energy bands, which are the result of
the combined analysis of the light curves at all wavelengths. All
other parameters have been marginalized. The black contours show the
results for a logarithmic prior on the source area, and the gray
contours are for a linear prior. With a logarithmic prior we cannot
put a lower limit on the \ac{UV} or X-ray size, but the distribution
increases with decreasing area until it hits the resolution limit of
our magnification patterns, depicted as a shaded region in
Figure~\ref{fig:massarea}. In the top panels of Figure~\ref{fig:area}
we plot the projection of this joint distribution along the vertical
axis, marginalizing over $\langle M \rangle$. The effect of the
resolution limit can be seen on the logarithmic prior (black) curves,
causing an artificial decrease in probability at small source
areas. In the bottom panels of Figure~\ref{fig:area} we avoid this
problem by taking a slice across the joint distribution at fixed
$\langle M \rangle = 0.3M_\odot$. With a logarithmic prior, we find
that in the $R$ band $\log(A/\mathrm{cm}^2) = 30.84\pm0.60$
(68\% confidence), while in the soft and hard X-ray bands,
respectively, $\log(A/\mathrm{cm}^2) < 29.89$ and $30.07$ at 95\%
confidence. Because our resolution limit prevents the X-ray
distributions from converging, the confidence level is probably higher
than 90\%. The \ac{UV} area has a very broad distribution which
decreases more or less monotonically from the resolution limit out to
larger areas, with a tail extending to larger areas than even the
optical distribution. Since the linear prior puts a much greater
weight on larger sources, these distributions are biased high relative
to the logarithmic prior, especially in the \ac{UV}, where the area
distribution is wide. With this prior, the logarithm of the source
areas in the $R$ band, the soft X-ray band, and the hard X-ray band
are $31.53\err{0.41}{0.53}$, $29.86\err{0.62}{0.72}$, and
$30.07\err{0.58}{0.73}$, respectively. Since the \ac{UV} distributions
are so broad and so prior-dependent, it is probably best to view the
\ac{UV} result as inconclusive. In Figures~\ref{fig:massarea} and
\ref{fig:area} we plot the square of the black hole's gravitational
radius as a vertical line, using the \citet{Peng:2006p616} mass of
$5\times10^8 M_\odot$. We can think of this as a hard lower limit on
the area, keeping in mind that the black hole mass is uncertain by a
factor of 3 or so. Aside from this we cannot set a lower limit on the
\ac{UV} or X-ray size without resorting to the linear prior.

\ifx \emulmacro \undefined
\else
\begin{figure*}
  \includegraphics[width=\textwidth]{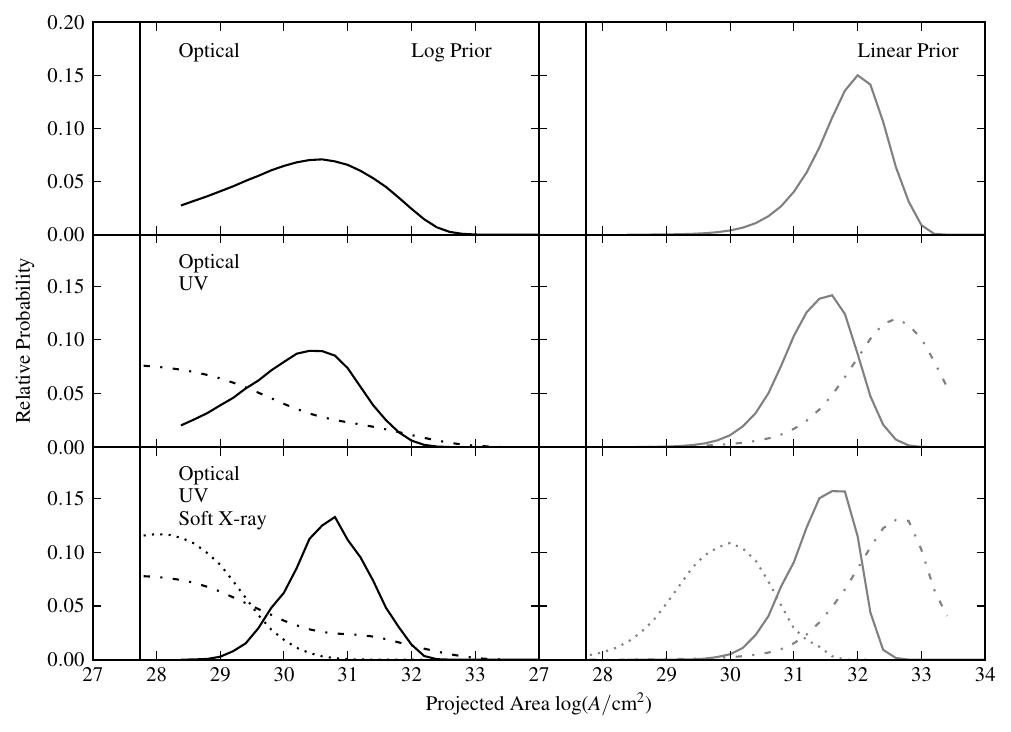}
  \caption{Posterior probability distributions for the projected area
  of the source for fixed $\langle M \rangle = 0.3M_\odot$,constrained
  by various combinations of data. Top panels: optical source area,
  constrained only by the $R$-band light curve. Middle panels: optical
  (solid) and UV (dot-dashed) source areas, constrained by $R$-band
  and UV light curves. Bottom panels: optical, UV, and soft X-ray
  (dotted) source areas, constrained by the $R$-band, UV, and soft
  X-ray light curves. The left (right) panels use a logarithmic
  (linear) prior. The square of the black hole's gravitational radius
  is shown with a vertical line. Compare to the bottom panels of
  \protect{Figure~\ref{fig:area}}, where the source areas are
  constrained by all four light curves.}
  \label{fig:area2}
\end{figure*}

\fi

It is interesting to ask which combinations of wavelengths have the
greatest power to constrain the area. Obviously, without the \ac{UV}
light curves we have no information about the corresponding source
size, and likewise in X-rays. Nor have we simulated either the hard or
soft X-ray or \ac{UV} in the absence of the $R$-band light curve. But
what improvement on the optical size is achievable by adding, e.g.,
just the \ac{UV} data?  Figure~\ref{fig:area2} shows the source area
probability distributions obtained from each step in our sequence of
simulations. The top panels show the constraint on the $R$-band area
resulting from modeling only the optical light curve. With a
logarithmic prior, it is broad enough that it does not converge at
small areas. Adding additional constraints from the \ac{UV} (middle
panels) and soft X-ray (bottom panels) light curves cause the optical
size distribution to become narrower, particularly in the X-ray
case. Comparing this figure to the lower panels of
Figure~\ref{fig:area}, it is clear that the addition of the X-ray data
has significant power to exclude small-area solutions for the optical
size. This is to be expected given the greater variability seen in the
X-ray band than at other wavelengths (see Figure~\ref{fig:lc}). The
high X-ray variability rules out low-velocity solutions which are the
only way that a very small optical source can exhibit (relatively)
small amplitudes of variability. The X-ray and \ac{UV} data have
relatively modest effects on each other's distributions.

\subsection{Disk Inclination}

The inclination of the accretion disk and the position angle of its
major axis have subtle effects on the microlensing light curves, and
can only be realistically simulated with dynamic magnification
patterns, as is done for \qtwotwolong\ by
\citet{Poindexter:2010p668}. The microlensing variability of \hefour\
is not as prominent as that of \qtwotwo, and its lensing geometry is
such that the random stellar motions are less important relative to
the motion of the source across the patterns. For these reasons, it is
not surprising that our posterior probability distributions show no
significant preference for particular values of the inclination $\cos
i$ and the major axis position angle $\phi_a$. Therefore, we treat
these parameters as undetermined, and simply marginalize the
likelihood over them.

\ifx \emulmacro \undefined
\else
\begin{figure}
  \includegraphics[width=\columnwidth]{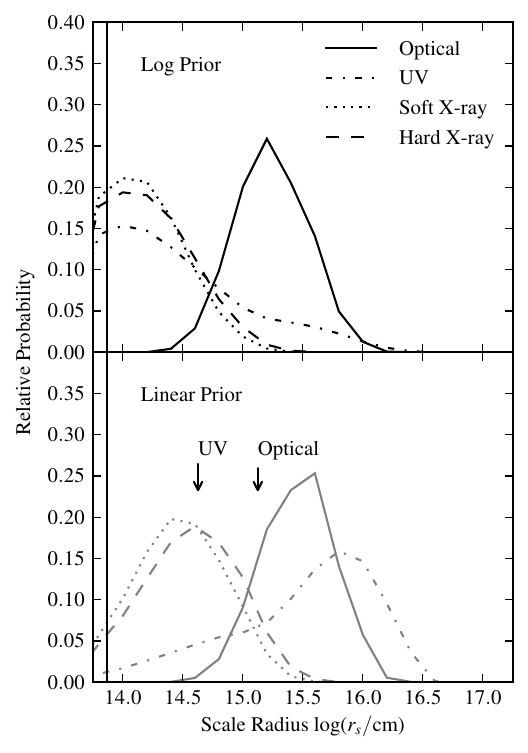}
  \caption{Posterior probability distributions for the scale radius of
  the source at a variety of wavelengths. The top (bottom) panel uses
  a logarithmic (linear) prior. The arrows mark the observed-frame UV
  and $R$-band scale radii expected from thin disk theory; the flux
  estimates are smaller by 0.38 dex. The vertical line on the left
  marks the estimated gravitational radius of the black hole. The mean
  mass $\langle M \rangle$ is fixed at $0.3M_\odot$, and all other
  parameters are marginalized.}
  \label{fig:rad}
\end{figure}

\begin{figure}
  \includegraphics[width=\columnwidth]{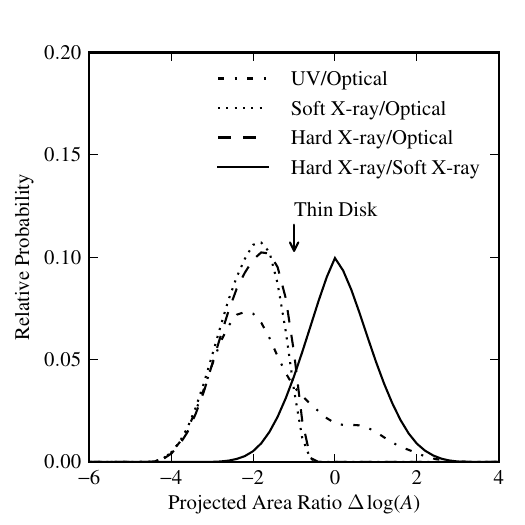}
  \caption{Posterior probability distributions for the ratios of
  projected areas, expressed as logarithmic differences. The arrow
  denotes the ratio of UV area to optical area predicted by the
  standard thin disk theory; compare it to the dot-dashed curve. The
  mean mass $\langle M \rangle$ is fixed at $0.3M_\odot$, and all
  other parameters are marginalized. We use a logarithmic prior.}
  \label{fig:arearat}
\end{figure}

\begin{figure}
  \includegraphics[width=\columnwidth]{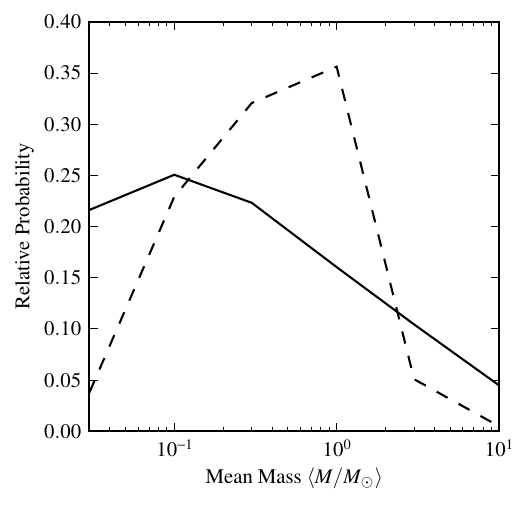}
  \caption{Posterior probability distribution for the mean mass
  $\langle M \rangle$ of the microlens stars. All other parameters are
  marginalized. This result is taken from the optical-only simulation,
  to avoid biases arising from the lack of convergence at small X-ray
  and UV sizes; see text for details. The solid line uses a flat
  logarithmic prior, while the dashed line uses the result of
  \protect{\citet{Poindexter:2010p658}} as a prior.}
  \label{fig:mass}
\end{figure}

\fi

\subsection{Disk Scale Radius}

Since $A$ is its projected area of the accretion disk, it depends both
on the physical size of the disk and its inclination, in the sense
that more highly inclined disks must be larger to present the same
area. When calculating the Bayesian integrals, we construct posterior
probability functions for the disk's scale radius $r_s(\lambda)$,
removing the inclination effects. This parameter is the radius in the
thin disk where $k_B T_\mathrm{eff}(r) = h c/\lambda$. Assuming the
accretion disk temperature slope $\beta = 3/4$ as in the thin disk
model, it is related to the half-light radius of the accretion disk by
the factor $r_{1/2}/r_s = 2.44$. The posterior probability
distributions for $r_s$ are shown in Figure~\ref{fig:rad} for fixed
$\langle M \rangle = 0.3 M_\odot$.  With a logarithmic prior on the
scale radii, we find an $R$-band size $\log(r_s/\mathrm{cm}) =
15.23\err{0.34}{0.33}$. By way of comparison, for the adopted black
hole mass of $5\times10^{8} M_\odot$, an Eddington ratio of $0.3$, and
an accretion efficiency of $0.1$, at this wavelength thin disk theory
predicts a value of $15.12$. A flux-based estimate yields a value of
$14.74$, assuming $\beta=3/4$. (See, e.g., \citet{Mosquera:2011p96}
for a description of these two estimators.)  In the \ac{UV} we can
place a 95\% confidence upper limit on the size $\log(r_s/\mathrm{cm})
< 15.66$, consistent with both the theory size of $14.63$ and the flux
size of $14.24$. The expected theory sizes for the optical and \ac{UV}
disk are shown in Figure~\ref{fig:rad}. The soft and hard X-ray radii
have probability distributions similar to each other, indicating sizes
smaller than $14.80$ and $14.90$ at 95\% confidence,
respectively. With the logarithmic prior, the \ac{UV} and X-ray
distributions all show a small decline at the smallest radii, but this
is a resolution artifact similar to the one shown in
Figure~\ref{fig:area}. Namely, at the smallest scale radii only
face-on disks are resolved, so only trials with $\cos i \approx 1$
contribute. This artifact is only important for the two leftmost
points in Figure~\ref{fig:rad}. Our logarithmic-prior estimate for the
$r$ band disk half-light radius, $\log(r_{1/2}/\mathrm{cm}) =
\log(2.44r_s/\mathrm{cm}) = 15.62\err{0.34}{0.33}$, is slightly
smaller than, but still consistent with previous estimates at similar
wavelengths. \citet{Morgan:2010p1129}, \citet{Blackburne:2011p34} and
\citet{Mosquera:2011p145} respectively find values of
$16.1\err{0.5}{0.7}$, $16.04\pm0.38$, and $16.32\err{0.27}{0.84}$. In
comparing these results, we note that our present calculations account
for more sources of systematic uncertainty than these other studies,
which is why they have comparable uncertainties despite using superior
data.

\subsection{Wavelength Dependence of Disk Area}

Our model parametrizes the dependence of the disk area on wavelength
using the area at each wavelength, rather than a normalizing area and
power-law slope. Since the X-rays do not originate from thermal thin
disk emission, only the optical and \ac{UV} sizes constrain the
wavelength slope. Therefore the two approaches are equivalent in our
case, with two parameters each. To estimate the dependence of disk
area on wavelength, we calculate posterior probability distributions
for the ratio of areas. Figure~\ref{fig:arearat} shows distributions
for the logarithm of the \ac{UV} area to optical area ratio, along
with similar distributions for soft X-ray to optical, hard X-ray to
optical, and hard X-ray to soft X-ray. These distributions use a
logarithmic prior only, as a linear prior does not make sense for such
ratios. We also mark the expected value of the \ac{UV} to optical
ratio for the thin disk case where $\Delta \log A = 8/3 \Delta \log
\lambda$. This value is consistent with our probability distribution,
which is not surprising given the width of the \ac{UV} area
distribution. Although it is not relevant for the thin disk model, the
ratio of hard X-ray area to soft X-ray area is interesting because of
the possibility of structure in the X-ray corona. We do not find
evidence of a difference in sizes: the distribution peaks just at
unity (though it is broad enough that significant differences in the
source area at different X-ray energies are not ruled out). This runs
a bit counter to the results of \citet{Chen:2011pL34}, who measure
X-ray light curves indicating energy-dependent size differences in
\qtwotwolong. More densely-sampled X-ray light curves would be
invaluable in addressing this question.

\subsection{Mean Microlens Mass}

Figure~\ref{fig:mass} shows the posterior probability distribution for
the mean mass of the stars $\langle M \rangle$. Like the source area,
this measurement is affected by the resolution limits of our
magnification patterns (see Figure~\ref{fig:massarea}). To avoid this
problem, which would bias our mass results toward smaller mean masses,
we use the results of the simulation that is constrained only by the
$R$-band light curve. Like \citet{Poindexter:2010p658}, we partially
break the degeneracy between velocity and mean mass by virtue of our
dynamic magnification patterns, but as we have mentioned, the small
random stellar velocities in the \hefour\ system reduce the power of
this method. Indeed, our posterior velocity distribution is
indistinguishable from the prior, indicating that we have not strongly
broken the degeneracy. Our mean mass probability distribution is
consistent with other estimates of the mean mass
\citep{Poindexter:2010p658}. Since our posterior probability
distribution is independent of that of \citet{Poindexter:2010p658}, we
multiply the two distributions together and plot them as a dashed
curve in Figure~\ref{fig:mass}. This is equivalent to using their
distribution as a prior.

\section{Conclusions}
\label{sec:conclusions}

Based on eight seasons of $R$-band monitoring data, nine epochs of
\ac{UV} photometry from \ac{HST}, and four epochs of X-ray fluxes in
two energy bands from \chandra, we have put limits on the projected
area and scale radius $r_s$ of the accretion disk of
\hefourlong. Using logarithmic priors, we find in the observed $R$
band that the scale radius $\log(r_s/\mathrm{cm})$ is
$15.23\err{0.34}{0.33}$, or about 23 gravitational radii assuming a
black hole mass $M_\mathrm{BH} = 5\times10^8 M_\odot$ \citep[based on
  a \textsc{Civ} line width measurement;][]{Peng:2006p616}. This disk
size is consistent with, though slightly smaller than, previous
results \citep{Morgan:2010p1129, Blackburne:2011p34,
  Mosquera:2011p145} It is about $0.1$ dex larger than the size
predicted by thin disk theory, and $0.5$ dex larger than the size
estimated from the optical flux. The tension with the flux size is
similar to other quasars with microlensing measurements, nearly all of
which are larger than their flux would suggest. The \ac{UV} size is
more poorly constrained, but we can say that it is most likely smaller
than the optical size. In soft ($0.4$\,--\,$1.3$\,keV) and hard
($1.3$\,--\,$8$\,keV) X-ray energy bands respectively, we can set
upper limits of $\log(r_s/\mathrm{cm}) = 14.80$ and $14.90$, or 9 and
11 gravitational radii, with at least 95\% confidence. We also compute
probability distributions for the area ratios between various
wavelengths. With a very broad distribution, the \ac{UV}/optical ratio
is consistent with both thin disk theory and with previous estimates
of the wavelength slope, and the hard X-ray/soft X-ray area ratio is
most consistent with unity. Adopting a range of values for the mean
mass $\langle M \rangle$ of the stars in the lens galaxy, we find
results consistent with earlier studies, with a most likely range of
0.1\,--\,1\,$M_\odot$.

\acknowledgements

This research was supported in part by NSF grant AST-1009756. Support
for \textit{HST} programs \#11732 and \#12324 was provided by NASA
through a grant from the Space Telescope Science Institute, which is
operated by the Association of Universities for Research in Astronomy,
Inc., under NASA contract NAS5-26555. Further support for this work
was provided by the National Aeronautics and Space Administration
through Chandra Award Number 11121 issued by the Chandra X-ray
Observatory Center, which is operated by the Smithsonian Astrophysical
Observatory for and on behalf of the National Aeronautics Space
Administration under contract NAS8-03060. This work was supported in
part by an allocation of computing time from the Ohio Supercomputer
Center.

%%%%%%%%%%%%%%%%%%%%%%%%%%%%%%%%%%%%%%%%%%%%%%
\bibliography{ms}

%%%%%%%%%%%%%%%%%%%%%%%%%%%%%%%%%%%%%%%%%%%%%%
\ifx \emulmacro \undefined

\clearpage

\begin{deluxetable*}{ccccc}
\tablewidth{0pt}
\tablecaption{$R$-band Light Curves
  \label{tab:optlc}}
\tablehead{
  \colhead{$\mathrm{HJD}-2450000$} &
  \colhead{A} &
  \colhead{B} &
  \colhead{C} &
  \colhead{D}
} 
\startdata
$2863.883$ & $2.065\pm0.012$ & $2.632\pm0.019$ & $2.592\pm0.018$ & $2.821\pm0.022$ \\ 
$2871.829$ & $2.140\pm0.014$ & $2.628\pm0.022$ & $2.561\pm0.019$ & $2.790\pm0.023$ \\ 
$2877.843$ & $2.134\pm0.012$ & $2.644\pm0.017$ & $2.670\pm0.016$ & $2.803\pm0.018$ \\ 
$2884.784$ & $2.141\pm0.012$ & $2.653\pm0.017$ & $2.661\pm0.017$ & $2.870\pm0.019$ \\ 
$2891.827$ & $2.172\pm0.015$ & $2.689\pm0.023$ & $2.685\pm0.022$ & $2.831\pm0.024$ \\ 
$2899.843$ & $2.171\pm0.012$ & $2.728\pm0.018$ & $2.712\pm0.017$ & $2.856\pm0.019$ \\ 
$2906.848$ & $2.184\pm0.008$ & $2.729\pm0.011$ & $2.745\pm0.011$ & $2.891\pm0.012$ \\ 
$2916.803$ & $2.192\pm0.010$ & $2.720\pm0.015$ & $2.713\pm0.014$ & $2.953\pm0.016$ \\ 
$2919.828$ & $2.177\pm0.013$ & $2.759\pm0.020$ & $2.738\pm0.019$ & $2.910\pm0.021$ \\ 
$2926.780$ & $2.186\pm0.012$ & $2.736\pm0.018$ & $2.643\pm0.016$ & $2.885\pm0.019$
\enddata
\tablecomments{~Light curves are in uncalibrated
magnitudes. This table is published in its entirety online. A portion
is shown here for guidance regarding its form and content.}
\end{deluxetable*}

\begin{deluxetable*}{ccccc}
\tablewidth{0pt}
\tablecaption{Ultraviolet (F275W) Light Curves
  \label{tab:uvlc}}
\tablehead{
  \colhead{$\mathrm{HJD}-2450000$} &
  \colhead{A} &
  \colhead{B} &
  \colhead{C} &
  \colhead{D}
} 
\startdata
$5072.079$ & $18.566\pm0.002$ & $18.917\pm0.003$ & $18.867\pm0.003$ & $19.005\pm0.003$ \\
$5135.646$ & $18.664\pm0.003$ & $19.004\pm0.003$ & $18.983\pm0.003$ & $19.096\pm0.003$ \\
$5179.185$ & $18.830\pm0.004$ & $19.198\pm0.005$ & $19.146\pm0.005$ & $19.288\pm0.006$ \\
$5298.162$ & $18.780\pm0.003$ & $19.197\pm0.003$ & $19.090\pm0.003$ & $19.247\pm0.004$ \\
$5400.013$ & $18.684\pm0.003$ & $19.064\pm0.003$ & $18.953\pm0.003$ & $19.136\pm0.003$ \\
$5453.678$ & $18.889\pm0.003$ & $19.244\pm0.004$ & $19.204\pm0.003$ & $19.408\pm0.004$ \\
$5550.662$ & $18.992\pm0.003$ & $19.433\pm0.004$ & $19.219\pm0.003$ & $19.676\pm0.004$ \\
$5615.296$ & $18.878\pm0.003$ & $19.323\pm0.004$ & $19.104\pm0.003$ & $19.443\pm0.004$ \\
$5662.224$ & $18.742\pm0.003$ & $19.066\pm0.003$ & $18.949\pm0.003$ & $19.119\pm0.003$ \\
$5783.750$ & $18.450\pm0.003$ & $18.854\pm0.003$ & $18.642\pm0.003$ & $19.012\pm0.004$
\enddata
\tablecomments{~Light curves are in ST magnitudes.}
\end{deluxetable*}

\begin{deluxetable}{ccccccc}
\tablewidth{0pt}
\tablecaption{Lens Model Parameters
  \label{tab:modparam}}
\tablehead{
  \colhead{$f_{M/L}$} &
  \colhead{Img} &
  \colhead{$\kappa_\mathrm{tot}$} &
  \colhead{$\kappa_*$} &
  \colhead{$\gamma$} &
  \colhead{$\phi_\gamma$\tablenotemark{a}} &
  \colhead{$\mu$\tablenotemark{b}}
}
\startdata
0.1                  & A & 0.662 & 0.0151    & 0.223    & \phn63\fdg2 & $+15.6$ \\
                     & B & 0.789 & 0.0292    & 0.322    &    344\fdg7 & $-16.7$ \\
                     & C & 0.663 & 0.0151    & 0.226    &    270\fdg3 & $+16.1$ \\
                     & D & 0.839 & 0.0367    & 0.342    &    167\fdg0 & $-10.9$ \\
0.3                  & A & 0.547 & 0.0464    & 0.300    & \phn63\fdg0 & $+8.68$ \\
                     & B & 0.674 & 0.0886    & 0.469    &    344\fdg7 & $-8.79$ \\
                     & C & 0.548 & 0.0465    & 0.304    &    270\fdg5 & $+8.92$ \\
                     & D & 0.723 & 0.0109    & 0.511    &    167\fdg1 & $-5.43$ \\
0.95                 & A & 0.157 & 0.148\phn & 0.553    & \phn62\fdg6 & $+2.48$ \\ 
                     & B & 0.299 & 0.289\phn & 0.961    &    344\fdg7 & $-2.31$ \\ 
                     & C & 0.158 & 0.147\phn & 0.562    &    270\fdg8 & $+2.54$ \\ 
                     & D & 0.358 & 0.347\phn & 1.07\phn &    167\fdg0 & $-1.36$ 
\enddata
\tablenotetext{a}{Measured in degrees East of North.}
\tablenotetext{b}{Sign indicates parity. Magnifications calculated using
\mbox{\protect{$\mu =
[(1-\kappa_\mathrm{tot}+\gamma)(1-\kappa_\mathrm{tot}-\gamma)]^{-1}$}}.}
\end{deluxetable}

\clearpage

\begin{figure*}
  \includegraphics[width=\textwidth]{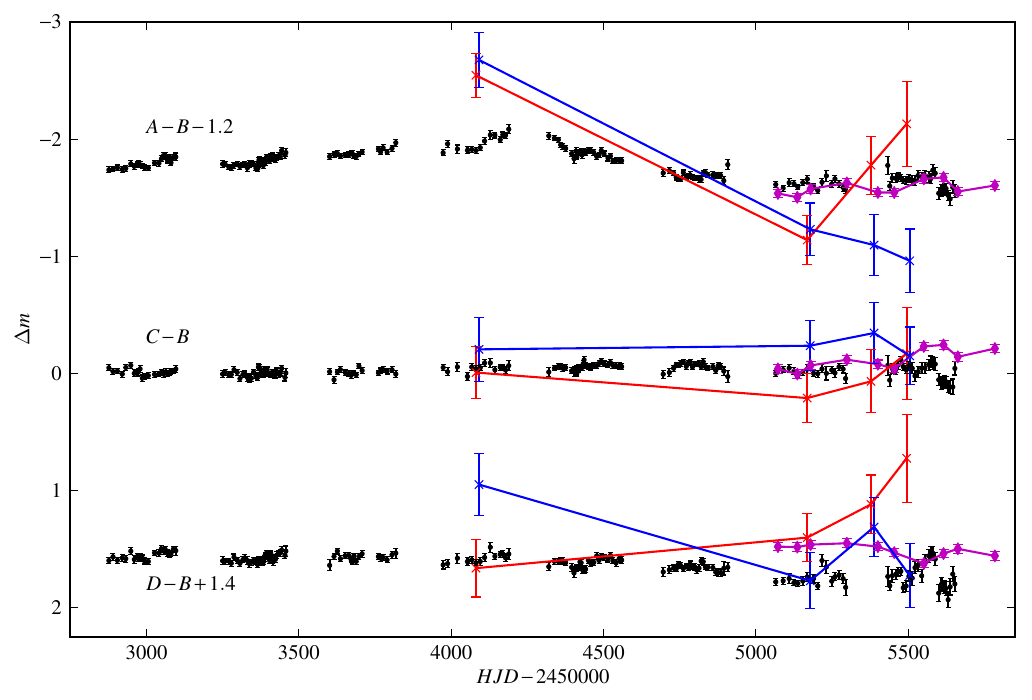}
  \caption{Light curves of the flux ratios between the images of
  \hefour, expressed in magnitudes. Black circles, magenta diamonds,
  red crosses, and blue crosses show the $R$-band, UV, soft X-ray, and
  hard X-ray data, respectively. The optical and UV curves have been
  corrected for the lensing time delays as described in
  \protect{Section~\ref{sec:data}}, and the optical light curve has
  been binned by a factor of two. Any remaining variability is
  considered to be due to microlensing. The light curves as shown are
  the input to the microlensing simulation. It is clear that the X-ray
  light curves have significantly more microlensing variability than
  the optical and UV curves.}
  \label{fig:lc}
\end{figure*}

\clearpage

\begin{figure*}
  \includegraphics[width=\textwidth]{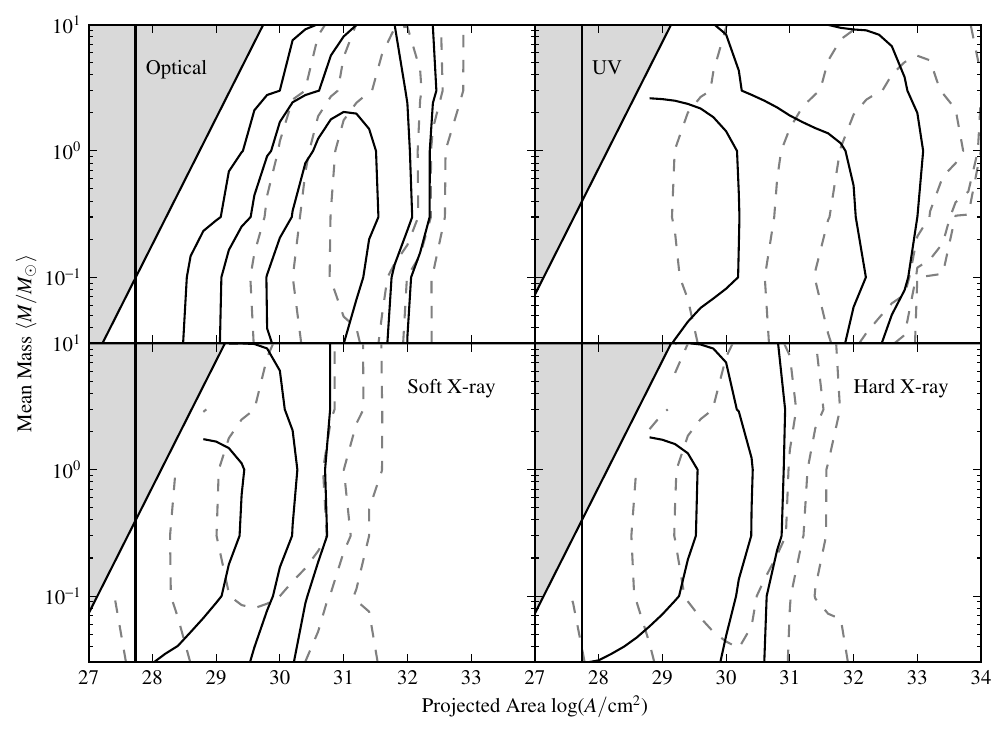}
  \caption{Joint posterior probability distributions for the projected
  area of the source and the microlens mean mass. The contours show
  39\%, 86\%, and 99\% enclosed probability (note the non-standard
  values). Black curves denote the logarithmic prior, and gray dashed
  curves the linear prior. The shaded zone indicates the region where
  the source area is smaller than the pixel size of the magnification
  patterns. The square of the black hole's gravitational radius (for
  $M_\mathrm{BH} = 5\times10^8M_\odot$) is shown with a vertical
  line. All other parameters are marginalized.}
  \label{fig:massarea}
\end{figure*}

\clearpage

\begin{figure*}
  \includegraphics[width=\textwidth]{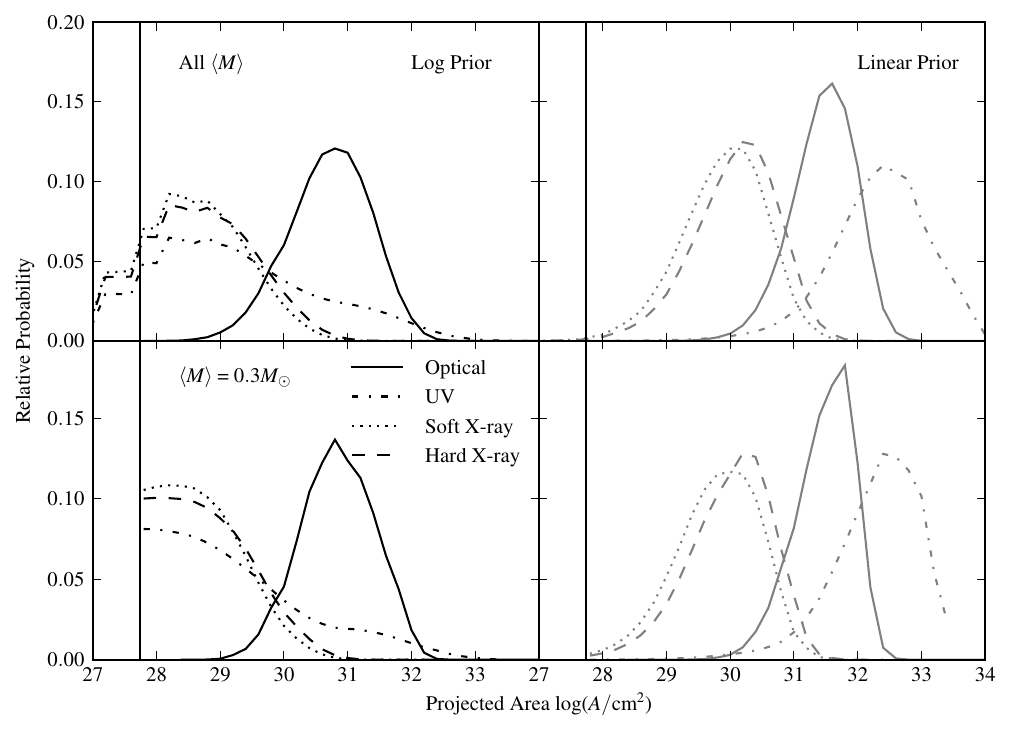}
  \caption{Posterior probability distributions for the projected area
  of the region producing the observed-frame $R$ band, UV, soft X-ray,
  and hard X-ray emission. The left (right) two panels use a
  logarithmic (linear) prior. In the top panels, all other variables
  have been marginalized, including $\langle M \rangle$, leading to an
  artificial decrease in probability at small areas (see
  \protect{Figure~\ref{fig:massarea}}). In the bottom panels, $\langle
  M \rangle$ is fixed at $0.3 M_\odot$. In all panels, the square of
  the black hole's gravitational radius is marked with a vertical
  line.}
  \label{fig:area}
\end{figure*}

\clearpage

\begin{figure*}
  \includegraphics[width=\textwidth]{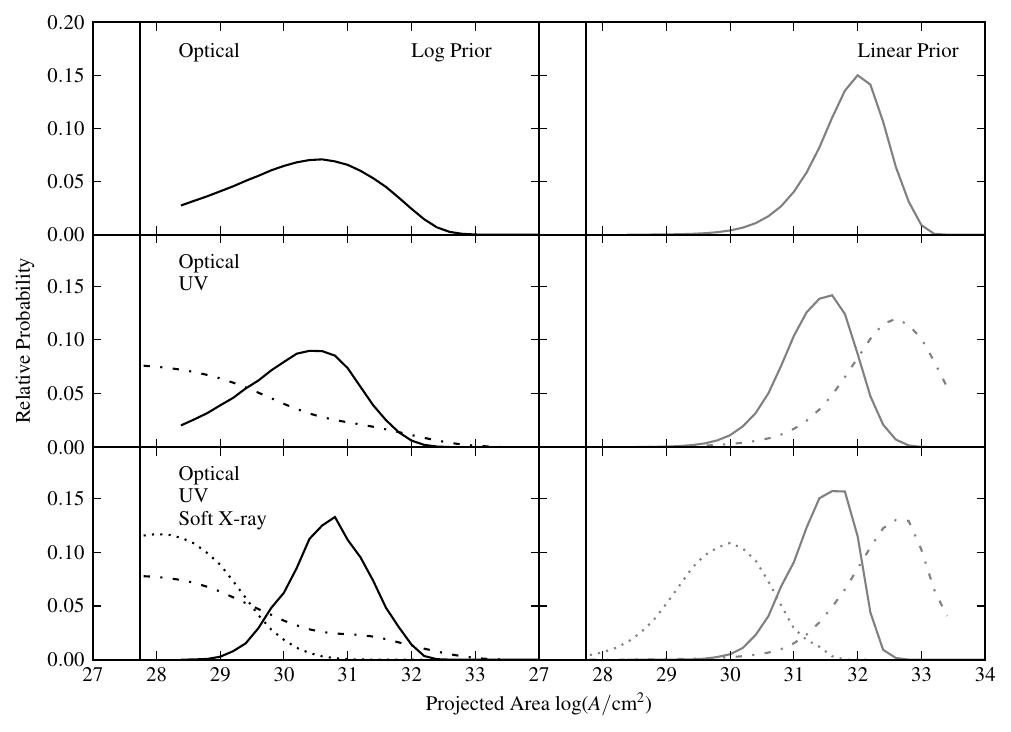}
  \caption{Posterior probability distributions for the projected area
  of the source for fixed $\langle M \rangle = 0.3M_\odot$,constrained
  by various combinations of data. Top panels: optical source area,
  constrained only by the $R$-band light curve. Middle panels: optical
  (solid) and UV (dot-dashed) source areas, constrained by $R$-band
  and UV light curves. Bottom panels: optical, UV, and soft X-ray
  (dotted) source areas, constrained by the $R$-band, UV, and soft
  X-ray light curves. The left (right) panels use a logarithmic
  (linear) prior. The square of the black hole's gravitational radius
  is shown with a vertical line. Compare to the bottom panels of
  \protect{Figure~\ref{fig:area}}, where the source areas are
  constrained by all four light curves.}
  \label{fig:area2}
\end{figure*}

\clearpage

\begin{figure}
  \includegraphics[width=\columnwidth]{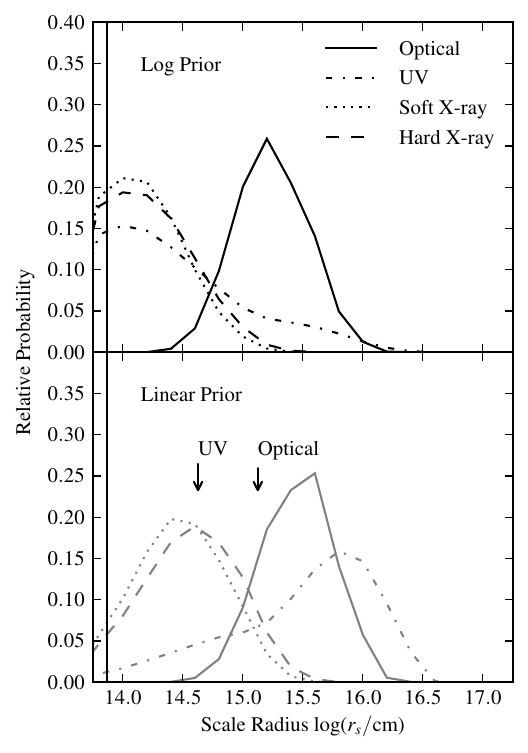}
  \caption{Posterior probability distributions for the scale radius of
  the source at a variety of wavelengths. The top (bottom) panel uses
  a logarithmic (linear) prior. The arrows mark the observed-frame UV
  and $R$-band scale radii expected from thin disk theory; the flux
  estimates are smaller by 0.38 dex. The vertical line on the left
  marks the estimated gravitational radius of the black hole. The mean
  mass $\langle M \rangle$ is fixed at $0.3M_\odot$, and all other
  parameters are marginalized.}
  \label{fig:rad}
\end{figure}

\clearpage

\begin{figure}
  \includegraphics[width=\columnwidth]{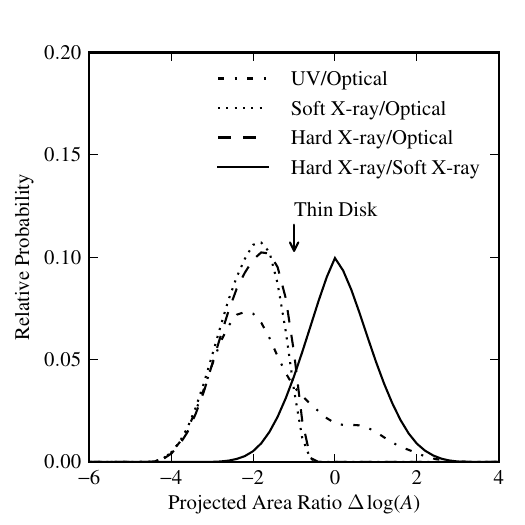}
  \caption{Posterior probability distributions for the ratios of
  projected areas, expressed as logarithmic differences. The arrow
  denotes the ratio of UV area to optical area predicted by the
  standard thin disk theory; compare it to the dot-dashed curve. The
  mean mass $\langle M \rangle$ is fixed at $0.3M_\odot$, and all
  other parameters are marginalized. We use a logarithmic prior.}
  \label{fig:arearat}
\end{figure}

\clearpage

\begin{figure}
  \includegraphics[width=\columnwidth]{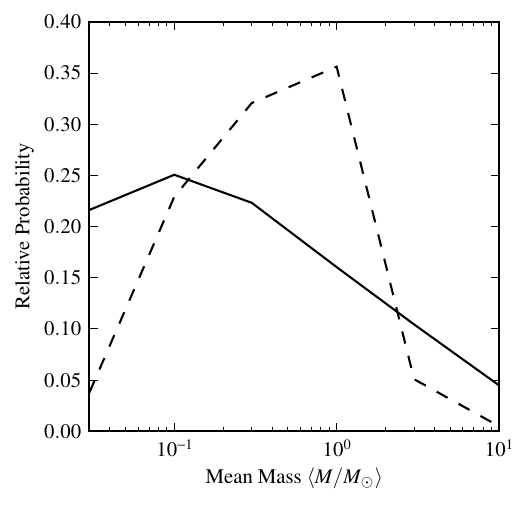}
  \caption{Posterior probability distribution for the mean mass
  $\langle M \rangle$ of the microlens stars. All other parameters are
  marginalized. This result is taken from the optical-only simulation,
  to avoid biases arising from the lack of convergence at small X-ray
  and UV sizes; see text for details. The solid line uses a flat
  logarithmic prior, while the dashed line uses the result of
  \protect{\citet{Poindexter:2010p658}} as a prior.}
  \label{fig:mass}
\end{figure}

\else
\fi

\end{document}